\documentclass{moriond}

\usepackage{amssymb}
\usepackage{amsmath}
\def\Journal#1#2#3#4{{#1} {\bf #2}, #3 (#4)}

\def\be{\begin{equation}}
\def\ee{\end{equation}}
\def\bea{\begin{eqnarray}}
\def\eea{\end{eqnarray}}

\newcommand{\Photo}{}

\begin{document}
\vspace*{4cm}
\title{Measurement of the running of the fine structure constant below 1 GeV with the KLOE detector}

\author{V. De Leo on behalf of the KLOE-2 Collaboration}

\address{INFN Sezione di Roma Tor Vergata}

\maketitle\abstracts{
The precision measurement of the $d\sigma(e^+e^- \rightarrow \mu^+\mu^- \gamma)/d\sqrt{s}$ cross section with the photon emitted
in the initial state with the KLOE detector has been used to measure the running of the QED coupling constant $\alpha(s)$ in the energy range 0.6 $< \sqrt{s} <$ 0.975 GeV in the time-like region. 
We were able to achieve a significance of the hadronic contribution to the running of $\alpha(s)$ of more than 5$\sigma$ with a clear contribution of the $\rho - \omega$ resonances to the photon
propagator. The real and imaginary part of the shift $\Delta\alpha$ to the running has been estracted and a fit of the real part  allowed us to measure the branching fraction $BR(\omega \rightarrow \mu^+\mu^-)$ = (6.6$\pm$1.4$_{stat}$$\pm$1.7$_{syst}$ )$\cdot$10$^{-5}$. }



\section{Introduction}

Tests of the Standard Model (SM) as well as establishing
possible new physics deviations from it require the very precise knowledge
of a set of input parameters like the fine structure constant $\alpha$, the Fermi
constant $G_\mu$ and the Z boson mass $M_Z$.
In QED the electromagnetic  coupling constant $\alpha(s)$ depends logarithmically on the energy scale due to the vacuum polarization
that causes a partial screening of the charge in the low energy limit (Thomson limit) while at higher energy the strenght of the electromagnetic interaction grows.
Thus, the classical charge has to be replaced by a ``running charge":  
 \begin{equation}
{e^{2} \rightarrow {e^2(q^2)}} = {{e^{2}} \over {1+(\Pi_\gamma^{'}(q^2) - \Pi_\gamma^{'}(0))}}
\label{rin_e}
\end{equation}
where, in the perturbation theory, the lowest order diagram which contributes to $\Pi_\gamma^{'}(q^2)$ is the vacuum polarization diagram which describes the virtual creation and reabsorption of fermion pairs:\,$ \gamma^* \rightarrow e^{+} e^{-}, \mu^{+} \mu^{-}, \tau^{+} \tau^{-}, u \bar{u}, d \bar{d},...\rightarrow \gamma^*  $
at the leading order \cite{fred}.
In terms of the QED coupling constant $\alpha={e^{2}}/{4\pi}$: 

\begin{equation}
{\alpha(q^{2})} = {{\alpha} \over {(1-\Delta \alpha)}};  \hspace{0.2cm} 
 \Delta \alpha = -Re(\Pi_\gamma^{'}(q^2)-\Pi_\gamma^{'}(0)).
\label{rin_e}
\end{equation}

The various contributions to the shift in the fine structure constant come from the leptons (lep=e,$\mu$ and $\tau$), the 5 light quarks (u,b,s,c and the corresponding hadrons =had) and from the top quark:
$\Delta \alpha = {\Delta \alpha}_{lep} + \Delta^{(5)}\alpha_{had} + \Delta \alpha_{top} + ...$ .
    
The experimental difficulties in the measurement of the running of the coupling constant are related to the evaluation  
of the hadronic contribution $\Delta\alpha_{had}$ because the low energy contributions of the five light quarks u,d,s,c, and b cannot be reliably calculated using perturbative quantum chromodynamics
(p-QCD) due to the non-perturbative behaviour of the strong interaction at low energies;   
perturbative QCD only allows
us to calculate the high energy tail of the hadronic (quark)
contributions. In the lower energy region the hadronic contribution can be evaluated through a dispersion relation over the measured $e^+e^- \rightarrow$ hadrons cross section.
%
%
%
 \;Therefore, it is clear that the dominant uncertainty in the evaluation of $\Delta\alpha$ is given by
the experimental data accuracy. \\
In the following the measurement of the running of the QED coupling constant in the range \\ 0.6 $< \sqrt{s} <$ 0.975 GeV in the time-like region will be reported together with the extraction, for the first time, of the real and imaginary part of $\Delta\alpha$.

\section{The KLOE detector}

Data corresponding to an integrated luminosity of 1.7 fb$^{-1}$ were collected by the KLOE detector at DA$\Phi$NE, the Frascati $e^+e^-$ collider,
which operates at a center of mass energy W = $m_\phi\sim$ 1020 MeV .
The KLOE detector consists
of a large cylindrical drift chamber (DC), surrounded by a fine sampling lead-scintillating
fibers electromagnetic calorimeter (EMC) inserted in a 0.52 T magnetic field.
The DC \cite{Adinolfi1}, 4 m diameter and 3.3 m long, has full stereo geometry and operates with a
gas mixture of 90\% helium and 10\% isobutane. Momentum resolution is $\sigma(p_\perp)/p_\perp \leq$ 0.4\%.
Position resolution in r - $\phi$ is 150 $\mu m$ and $\sigma_z \sim 2 mm$. Charged tracks vertices are
reconstructed with an accuracy of $\sim$ 3 mm.
The EMC \cite{Adinolfi2} is divided into a barrel and two endcaps, for a total of 88 modules
and covers 98\% of the solid angle.
Cells close in time and space are grouped into a calorimeter cluster. The cluster
energy E is the sum of the cell energies, while the cluster time t and its position \textbf{r} are
energy weighted averages. The respective resolutions are $\sigma_E/E = 5.7\%/\sqrt {E (GeV)}$ and
$\sigma_t = 57 ps/\sqrt{E\,(GeV)}$  $\oplus$ 100 ps.

\section{Measurement of the running of $\alpha$}
The running of  $\alpha(s)$ has been obtained from the ratio between the precise measurement of the Initial State Radiation (ISR) process $e^+e^-\rightarrow \mu^+\mu^-\gamma$ and the Monte Carlo (MC) simulation without the vacuum polarization (VP) contribution, in other words, setting $\alpha(s)=\alpha(0)$:      
\begin{equation}
\lvert \frac{\alpha(s)}{\alpha(0)}\rvert ^2= \frac{d\sigma_{data} (e^+e^- \rightarrow \mu^+\mu^- \gamma(\gamma))\vert_{ISR}/d\sqrt{s}}{d\sigma^{0}_{MC}(e^+e^- \rightarrow \mu^+\mu^- \gamma(\gamma))\vert_{ISR}/d\sqrt{s}} .
\label{our_method}
\end{equation}

\begin{figure}[h]
\begin{center}
\includegraphics[width=15.0pc]{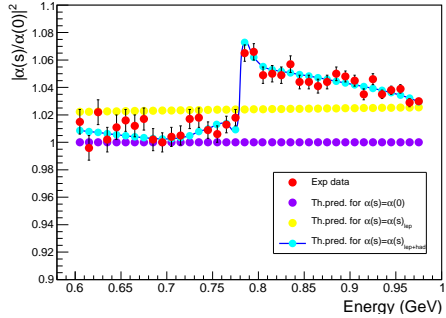}
\caption{The square of the modulus of the running $\alpha(s)$ in units of $\alpha(0)$ compared with the 
prediction (provided by the {\tt alphaQED} package) 
as a function of the dimuon invariant mass. The red points are the 
KLOE data with statistical errors; the violet points are the theoretical prediction for a fixed coupling ($\alpha(s)$ = $\alpha(0)$);
the yellow points are the prediction with only virtual lepton pairs contributing to the shift $\Delta\alpha(s)$ = $\Delta\alpha(s)_{lep}$, 
and finally the points with the solid line are the full QED prediction with both lepton and quark pairs contributing to the shift $\Delta\alpha(s)$ = $\Delta\alpha(s)_{lep+had}$.
}
\label{vp}
\end{center}
\end{figure}

The sample of $\mu\mu\gamma$ events is selected requiring a photon and two tracks of opposite curvature; the photon is emitted
 at small angle (SA),{\it i.e.} within a cone of $\theta_\gamma < 15^\circ$ around the beamline and the two charged muons are emitted at large polar angle, $50^\circ<\theta_\mu<130^\circ$ \cite{mio}. \\           
The experimental ISR $\mu^+\mu^-\gamma$ cross section is obtained from the observed number of $\mu\mu\gamma$ events ($N_{obs}$) and the background estimate ($N_{bckg}$) as: 

\begin{equation}
\frac{d\sigma(e^+e^-\rightarrow \mu^+\mu^-\gamma(\gamma))}{d\sqrt{s}}\biggr\rvert_{ISR}= \frac{N_{obs}-N_{bkg}}{\Delta\sqrt{s}}\cdot\frac{(1-\delta_{FSR}) }{\epsilon(\sqrt{s})\cdot \textit{L}},
\label{mmg}
\end{equation}

where $(1-\delta_{FSR})$ is the correction applied to remove the Final State Radiaton (FSR) contribution, 
 $\epsilon$ is global the efficiency, and $L$ is the integrated luminosity. \\
To separate the electrons from the pions or muons we used a particle identification estimator (L$\pm$), based on a pseudo-likelihood function using time-of-flight and calorimeter information (size and shape of the energy deposit). 
The muons were distiguished from the pions essentially by means of two selection cuts: the first one on the  $M_{TRK}$ ($M_{TRK}$ $<$115 MeV) that is a variable  computed requiring the energy and momentum conservation and the second on the $\sigma_{MTRK}$ that is constructed event by event with the error matrix of the 
fitted tracks at the point of closest approach (PCA). 
Cutting the high values of   $\sigma_{MTRK}$ the bad reconstructed tracks are rejected allowing a reduction of the $\pi\pi\gamma$ events contamination. The residual background is estimated  by fitting the observed   $M_{TRK}$ spectrum with a superposition
of MC simulation distributions describing signal
and $\pi^+\pi^-\gamma$, $\pi^+\pi^-\pi^0$ and $e^+e^-\gamma$ events. Additional background from the $e^+e^-\rightarrow e^+e^-\mu^+\mu^-$ process has been evaluated using the NEXTCALIBUR MC generator. The maximum contribution is 0.7\% at $\sqrt{s}$=0.6\,GeV. 
The contribution from $e^+e^-\rightarrow e^+e^-\pi^+\pi^-$ has been evaluated with the EKHARA generator and found to be negligible \cite{mio}.  
The measured $\mu^+\mu^-\gamma$ cross-section with only ISR is then compared  
 with the corresponding NLO QED calculation from PHOKHARA generator including the VP effects. The agreement between the two cross sections is excellent; the average ratio, using only the statistical errors, is 1.0006$\pm$0.0008.\\
%
By setting in the MC $\alpha(s)$ = $\alpha(0)$, the hadronic contribution to the photon propagator,
 with its
characteristic
 $\rho-\omega$ interference structure, is clearly visible in the data to MC ratio, as shown in Fig.~\ref{vp}. 

\section{Extraction of Real and Imaginary part of $\Delta\alpha$ and fit of Re$\Delta\alpha$}
\begin{figure}[h!]
\begin{center}
\includegraphics[width=14.0pc]{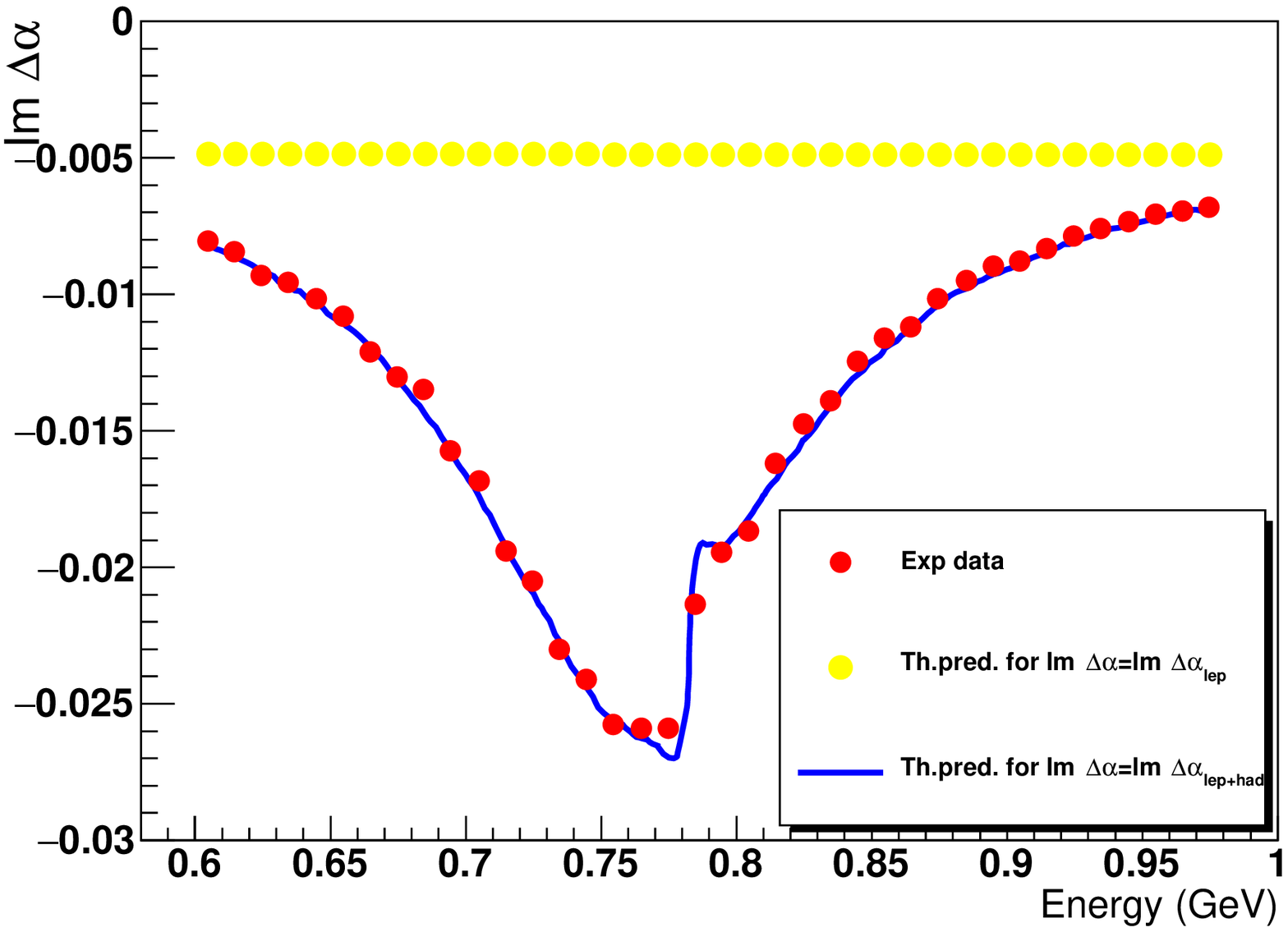}
\includegraphics[width=14.0pc]{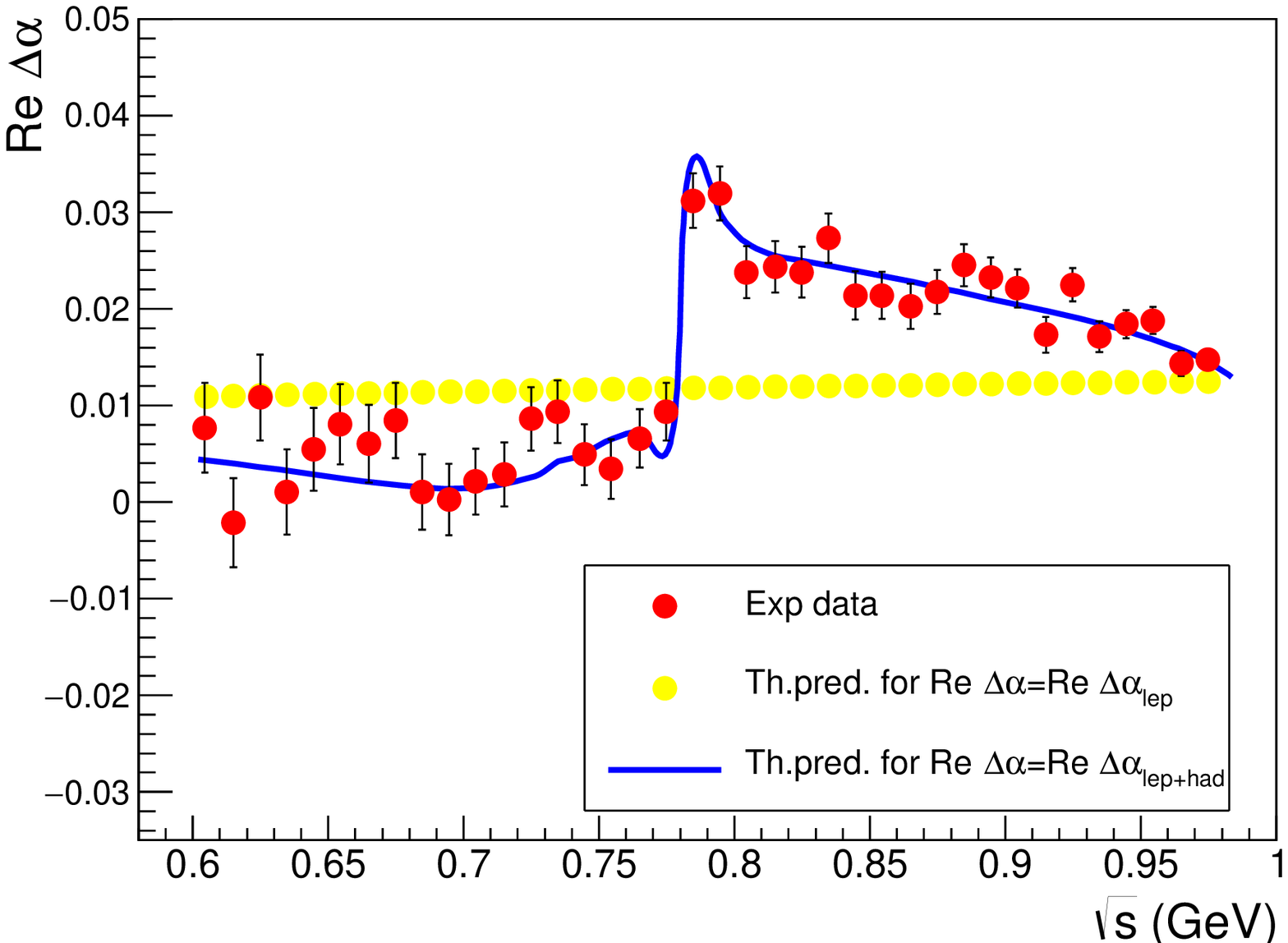}
\caption{Left: $\rm Im\,\Delta \alpha$ extracted from the KLOE data compared with the values provided by {\tt alphaQED} routine (without the KLOE data) for $\rm Im\,\Delta \alpha=\rm Im\,\Delta \alpha_{lep}$ (yellow points) and $\rm Im\,\Delta \alpha=\rm Im\,\Delta \alpha_{lep+had}$ only for $\pi\pi$ channels (blue solid line). Right: 
$\rm Re\,\Delta \alpha$ extracted from the experimental data with only the statistical error included compared with the {\tt alphaQED} prediction (without the KLOE data) 
 when $\rm Re\,\Delta \alpha=\rm Re\,\Delta \alpha_{lep}$ (yellow points) and $\rm Re\,\Delta \alpha=\rm Re\,\Delta \alpha_{lep+had}$ (blue solid line). 
}
\label{re_delta_alpha_all}
\end{center}
\end{figure}

Since the VP function $\Pi(q^2)$ is complex, both $\Delta\alpha$ and $\alpha(q^2)$ are complex quantities. Although usually the real part of $\Pi(q^2)$  is considered, which makes the effective coupling $\alpha(q^2)$ real, this approximation is not sufficient in presence of resonances, like the $\rho$.  
In this case the imaginary part become non-negligible and should be taken into account.
To evaluate the real part of $\Delta\alpha$ we used this simple relation: 
\begin{equation}
\rm Re\,\Delta \alpha = 1-\sqrt{\lvert \alpha(0)/\alpha(s) \rvert^2-(\rm Im\,\Delta \alpha)^2}
\label{re_delta_alpha}
\end{equation}  
defined in terms of the measured quantity $\lvert \alpha(s)/\alpha(0) \rvert^2$ and of the imaginary part that has been evaluated considering that for the optical theorem it can be related to the 
total cross section $\sigma(e^+e^-\to\gamma^*\to anything)$ (``anything'' means any possible state), where the precise relation reads~ \cite{fred}:
$Im \Delta \alpha = - \frac{\alpha}{3}\,R(s)$ 
with
$R(s) = \sigma_{tot}/\frac{4\pi\alpha(s)^2}{3s}$.
$R(s)$ takes into account the leptonic and hadronic contributions
$R(s)=R_{lep}(s)+R_{had}(s)$, where the leptonic part is given by: 
$R_{lep}(s)=\sqrt{1-\frac{4m_l^2}{s}} \left(1+\frac{2m_l^2}{s}\right),(l=e,\mu,\tau)$
while for the evaluation of the hadronic part 
we use 
only the 2$\pi$ hadronic contribution measured by KLOE \cite{Venanzo} which dominates in this region:
\begin{equation}
R_{had}(s)= \frac{1}{4} \left(1-\frac{4m_\pi^2}{s} \right)^\frac{3}{2} \vert F_\pi^0(s) \vert ^2
\end{equation}

where the Pion Form Factor must be deconvoluted by the VP effects: 
$\vert F_\pi^0(s)\vert ^2 = \vert F_\pi(s) \vert ^2\vert \frac{\alpha(0)}{\alpha(s)}\vert^2$.
The results obtained for the imaginary part of $\Delta\alpha(s)$ (Im $\Delta\alpha$)
are shown in left panel of the Fig.~\ref{re_delta_alpha_all} (the exp data are the red points) compared 
with the values given by the $R_{had}(s)$ compilaton of Ref. ~\cite{fj}  
 (blue solid line). The real part is shown on the right.    
  
The Re$\Delta\alpha$ has been fitted by a sum of the leptonic and hadronic contributions, where the hadronic contribution is
parametrized as a sum of $\rho(770)$, $\omega(782)$ and $\phi(1020)$ resonances components and a non resonant term
(param. by a first-order polynomial).

 For the $\omega$ and $\phi$ resonances
  \; a Breit-Wigner description was used \cite{mio}
\begin{equation}
Re\,\Delta \alpha_{V=\omega,\phi} =
\frac{3\sqrt{BR(V\to e^+e^-)\cdot BR(V\to\mu^+\mu^-)}}{\alpha M_V}\frac{s(s-M^2_V)\Gamma_V}{(s-M^2_V)^2+s\Gamma^2_V}
\end{equation}
where $M_V$ and $\Gamma_V$ are the mass and the total width of the mesons $V=\omega$ and $\phi$ while for the $\rho$ we use a Gounaris-Sakurai parametrization $BW^{GS}_{\rho(s)}$~ \cite{Gounaris:1968mw,Akhmetshin:2001ig} of the pion form factors, where we neglect the interference with the $\omega$, and the high excited states of the $\rho$ \cite{mio}.
%
Assuming lepton universality
and multiplying for the phase
space correction: 
$\xi =\Big(1+2\frac{m^2_\mu}{m^2_{\omega}}\Big)\Big(1-4\frac{m^2_\mu}{m^2_{\omega}}\Big)^{1/2}$ 
we found for the $BR(\omega\to\mu^+\mu^-)$ the following result:
$(6.6\pm1.4_{stat}\pm1.7_{syst})\cdot 10^{-5}$
compared to $(9.0\pm3.1)\cdot 10^{-5}$ from PDG \cite{mio}.

\section{Conclusions}
We present the first precision measurement of the running of $\alpha(s)$ in the energy region  0.6 $< \sqrt{s} <$ 0.975
and the strongest direct evidence  of the hadronic contribution to $\alpha(s)$ achieved in both time- and space-like regions
by a single experiment.
For the first time also the real and imaginary part of $\Delta\alpha$(s) have been
extracted showing clearly the importance of the role of the imaginary part.
By fitting the real part of $\Delta\alpha(s)$ and assumming the lepton universality, the
branching fraction $BR(\omega \rightarrow \mu^+\mu^-)$ = (6.6$\pm$1.4$_{stat}$$\pm$1.7$_{syst}$)$\cdot$10$^{-5}$ has also been
obtained.

\end{document}